\documentclass[letter]{aa}
\usepackage{graphicx}
\usepackage{txfonts}

\def\Lsun{L$_\odot$}
\def\Msun{M$_\odot$}

\def\Oiii{[O\,{\sc iii}]}

\def\Cii{[C\,{\sc ii}]}

\def\kms{km\,s$^{-1}$}
\def\zgt6{\hbox{{\it z}$\,>\,$6}}

\def\lsim{\mathrel{\rlap{\lower 3pt \hbox{$\sim$}} \raise 2.0pt \hbox{$<$}}}
\def\gsim{\mathrel{\rlap{\lower 3pt \hbox{$\sim$}} \raise 2.0pt \hbox{$>$}}}

\begin{document}

\authorrunning{Decarli et al.}
\titlerunning{A Mpc-scale quasar--galaxy association at cosmic dawn}

\title{Testing the paradigm: First spectroscopic evidence of a quasar--galaxy Mpc-scale association at cosmic dawn}
\author{
Roberto Decarli\inst{1}, Marco Mignoli\inst{1}, Roberto Gilli\inst{1}, Barbara Balmaverde\inst{2}, Marcella Brusa\inst{3,1}, Nico Cappelluti\inst{4}, Andrea Comastri\inst{1}, Riccardo Nanni\inst{1}, Alessandro Peca\inst{1}, Antonio Pensabene\inst{1,3}, Eros Vanzella\inst{1}, Cristian Vignali\inst{3,1}
}
\institute{
INAF -- Osservatorio di Astrofisica e Scienza dello Spazio di Bologna, via Gobetti 93/3, I-40129, Bologna, Italy. \email{ roberto.decarli@inaf.it} \and
INAF -- Osservatorio Astrofisico di Torino, Via Osservatorio 20, I-10025 Pino Torinese, Italy. \and
Dipartimento di Fisica e Astronomia, Universit\'{a} degli Studi di Bologna, Via P. Gobetti 93/2, I-40129, Bologna, Italy \and
Physics Department, University of Miami, Coral Gables, FL, 33124, USA}

\date{August 2019}
\abstract{State-of-the-art models of massive black hole formation postulate that quasars at \zgt6 reside in extreme peaks of the cosmic density structure in the early universe. Even so, direct observational evidence of these overdensities is elusive, especially on large scales ($\gg$1 Mpc) as the spectroscopic follow-up of \zgt6 galaxies is observationally expensive. Here we present Keck / DEIMOS optical and IRAM / NOEMA millimeter spectroscopy of a $z\sim6$ Lyman-break galaxy candidate originally discovered via broadband  selection, at a projected separation of 4.65 physical Mpc (13.94 arcmin) from the luminous $z$=6.308 quasar J1030+0524. This  well-studied field presents the strongest indication to date of a large-scale overdensity around a $z>6$ quasar. The Keck observations suggest a $z\sim6.3$ dropout identification of the galaxy. The NOEMA 1.2mm spectrum shows a 3.5$\sigma$ line that, if interpreted as \Cii{}, would place the galaxy at $z$=6.318 (i.e., at a line-of-sight separation of 3.9 comoving Mpc assuming that relative proper motion is negligible). The measured \Cii{} luminosity is $3\times10^8$\,\Lsun{}, in line with expectations for a galaxy with a star formation rate $\sim15$\,\Msun{}\,yr$^{-1}$, as inferred from the rest-frame UV photometry. Our combined observations place the galaxy at the same redshift as the quasar, thus strengthening the overdensity scenario for this \zgt6 quasar. This pilot experiment demonstrates the power of millimeter-wavelength observations in the characterization  of the environment of early quasars.}


\maketitle

\section{Introduction}

Since their discovery 20 years ago, quasars at {\it z}$\,\gsim\,$6 have shaped our understanding of early galaxy formation. Their immense luminosity is due to rapid gas accretion onto massive ($>10^{8}$\,\Msun) black holes \citep{derosa11,wu15}. Their host galaxies form stars at prodigious rates ($>500$\,\Msun{}\,yr$^{-1}$; see, e.g., \citealt{walter09, leipski14, venemans18}), supported by immense gaseous reservoirs \citep[e.g.,][]{bertoldi03, walter03, wang10, venemans17}. Models and numerical simulations of the formation of these early quasars almost unanimously agree that first quasars populated the extreme peaks of the cosmic matter distribution at those early cosmic times (age of the universe $<$ 1 Gyr; see, e.g., \citealt{begelman06, narayanan08, overzier09, bonoli09, bonoli14, angulo12, costa14}, but see also cautionary results from \citealt{fanidakis13} and \citealt{habouzit19}). In this scenario, the environment of \zgt6 quasars should present an excess in the number of companion galaxies, both on small ($<100$\,kpc) and large ($\gg1$\,Mpc) scales.

From an observational point of view, however, demonstrating the presence of these  overdensities has been challenging. On small scales studies are burdened by small samples and cosmic variance, on large scales by the need of covering wide areas, and in both cases by sensitivity limitations.  These factors lead to contrasting results in the literature (see, e.g., \citealt{overzier09, morselli14, mazzucchelli17} for discussions on the limitations of different approaches). On small scales the strongest evidence of overdensities around \zgt6 quasars comes from the discovery of \Cii{}-bright galaxies in the field of a few quasars observed with the Atacama Large Millimeter / sub-millimeter Array (ALMA) \citep{decarli17, trakhtenbrot17,willott17,neeleman19}. Recently, integral field observations with the MUSE instrument on the ESO Very Large Telescope also led to the discovery of close companions to \zgt6 quasars \citep{farina17}. However, these investigations  are all limited by the relatively small field of view of the available facilities.

To date, only six \zgt6 quasar fields have been studied on scales larger than 10 physical Mpc$^2$: SDSS J103027.10+052455.0 ($z$=6.308; hereafter, J1030+0524), SDSS J114816.65+525150.4 ($z$=6.419), SDSS J104845.07+463718.5 ($z$=6.228), and SDSS J141111.29+121737.4 ($z$=5.904) were observed with the Large Binocular Camera (LBC) on the Large Binocular Telescope (LBT) by \citet{morselli14};  CFHQ J232908.27-030158.8 ($z$=6.416) and VIK J030516.92-315055.9 ($z$=6.615) were observed using the Suprime-Cam on Subaru \citep{utsumi10, ota18}. Of these six fields, only two show indications of a clear overdensity of color-selected galaxies associated with the quasars: CFHQ J232908.27-030158.8 and J1030+0524. The latter field has been part of an extensive follow-up campaign using the Advanced Camera for Surveys \citep[][who already reported evidence of a galaxy overdensity around the quasar, despite the relatively small size of the field of view]{kim09} and the Wide Field Camera 3 on the {\em Hubble Space Telescope}, along with the LBC camera at the LBT, the Wide-field InfraRed Camera at the CFHT, {\em Spitzer}/InfraRed Array Camera and Multiband Imager Photometer, the Very Large Array at 1.4\,GHz, and {\em Chandra}. The field was also part of the Multiwavelength Survey by Yale-Chile \citep[MUSYC;][]{gawiser06} that provides additional imaging in UBVRIzJHK \citep{quadri07,blanc08}. The wealth of data on this field makes the case of a galaxy overdensity around J1030+0524 particularly compelling, among the \zgt6 quasars studied so far. 

\citet{balmaverde17} exploited the full multi-color dataset in the J1030+0524 field to produce a robust list of \hbox{{\it z}$\sim$6} galaxy candidates down to very faint $z_{\rm AB}$ magnitudes ($\approx$25.5). After measuring photometric redshifts, and improving the rejection of contaminants, 
they reinforced the case for a large-scale overdensity.
The field was followed up using optical spectrographs  at the Very Large Telescope and the Keck telescopes (Mignoli et al.\ in prep.), and the photometric \hbox{{\it z}$\sim$6} candidates were included whenever possible.

Here we present dedicated follow-up optical and millimeter spectroscopic observations of a color-selected dropout galaxy, ID22914, located at R.A.=10:30:00.90, Dec=+05:31:14.3 (i.e., at an angular separation of 13.94 arcmin from the quasar), corresponding to 4.65 physical Mpc, or 33.96 comoving Mpc in the adopted cosmology model. From the observed LBC $z$-band magnitude, $25.45\pm0.22$, we infer a UV-based star formation rate (SFR) of $\approx$15$\pm$3\,\Msun{}\,yr$^{-1}$, based on the scaling presented in \citet{kennicutt12}.
With a color {\it i}$-${\it z}$\,>\,$1.39$\pm$0.22 and a $z$-band magnitude fainter than 25.2\,mag, ID22914 did not satisfy the stringent color criterion applied by \citet{balmaverde17}. Nevertheless, we included it in the slit mask since, to exploit the  spectrograph's multiplex capability, we targeted all the useful {\it i}-dropout candidates.

Recently, the combination of optical and near-infrared spectroscopy of Ly$\alpha$ and the rest-frame UV emission of galaxies, and spectral scans at millimeter and sub-millimeter wavelengths targeting either \Cii{} 158\,$\mu$m or \Oiii{} 88\,$\mu$m, has allowed astronomers to push the redshift frontier in terms of spectroscopically confirmed galaxies \citep[e.g.,][]{inoue16,bradac17,laporte17,hashimoto18,hashimoto19,tamura19}. The new optical observations presented here on ID22914 were collected with the DEep Imaging Multi-Object Spectrograph (DEIMOS) at the Nasmyth focus of the 10m Keck II telescope \citep{faber03}. The 1.2\,mm data were secured with the IRAM / NOrthern Extended Millimeter Array (NOEMA). The combination of optical and millimeter spectroscopy allows us to search for a redshift determination via the hydrogen Ly$\alpha$ 1216\,\AA{} line and the singly-ionized carbon \Cii{} 158$\mu$m line. These are the brightest emission lines in their respective bands, and are commonly used as workhorses for redshift determination of high-redshift galaxies. 

Throughout the paper we adopt a concordance cosmology model with $H_0=70$\,\kms{}\,Mpc$^{-1}$, $\Omega_{\rm m}$=0.3, and $\Omega_\Lambda$=0.7, in agreement with the values measured by \citet{planck16}. In this framework, at the redshift of J1030+0524 \citep[$z$=6.3080, see][]{kurk07}, the luminosity distance is 61199 Mpc, and 1 arcsec corresponds to 5.56 kpc. All quoted magnitudes are in the AB photometric system. In the derivation of SFRs, we implicitly assume a \citet{chabrier03} stellar initial mass function.

%

\section{Observations and data reduction}\label{sec_obs}

\subsection{Keck observations}

On February 27, 2017, we used DEIMOS \citep{faber03} on the Keck~II telescope to obtain an optical spectrum of ID22914. The total exposure time for these observations was 4~hours. Aiming to cover rest-frame Lyman-$\alpha$ regions for the wide redshift range 4.50--7.0, we used the red-efficient 830 lines/mm grating and the OG550 order cut filter with the central wavelength of 8700\,\AA, which provide a spectral coverage between 7000\,\AA\ and 10400\,\AA.  The slit width was 1.2\arcsec, which gives a spectral resolution of $\approx$4\,\AA.
The data reduction was performed with standard \textsf{IRAF} routines for bias subtraction, flat-fielding, wavelength calibration, and optimal background subtraction. The standard star HD~93521 was used for the flux calibration. The seeing was excellent ($<$1\arcsec) so the slit losses were negligible,  confirmed by the fair agreement of the measured spectral flux above 8500\,\AA\ with the observed $z_{\rm AB}$ magnitude of ID22914.

\begin{figure}
\begin{center}
 \includegraphics[width=0.49\textwidth]{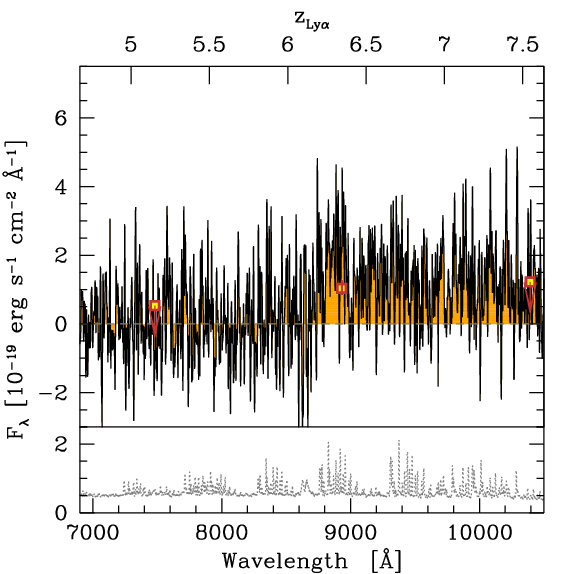}
\end{center}
 \caption{Optical Keck / DEIMOS spectrum of ID22914, the target of our study. In the bottom panel, we show the 1$\sigma$ error spectrum. Red squares indicate the i-, z-, and Y-band photometry. Limits refer to a 3$\sigma$ significance. The spectrum shows a blue continuum with a break around wavelength $\sim$8850\,\AA{}, implying a Ly$\alpha$ redshift of $z\approx6.3$.}
 \label{fig_lya}
\end{figure}

\begin{figure*}
\begin{center}
 \includegraphics[width=0.31\textwidth]{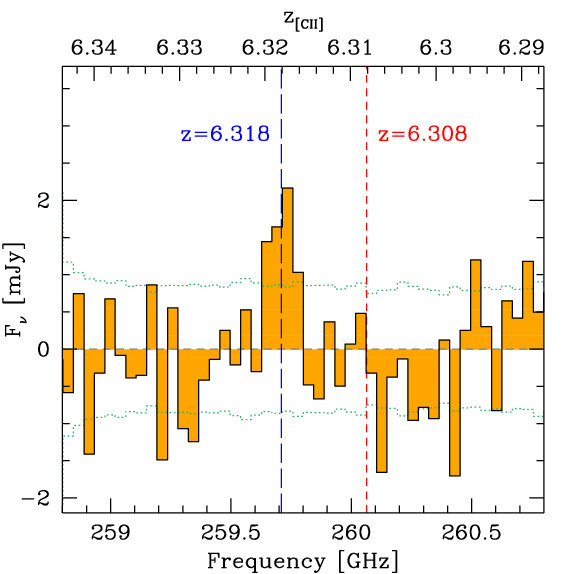}
 \includegraphics[width=0.33\textwidth]{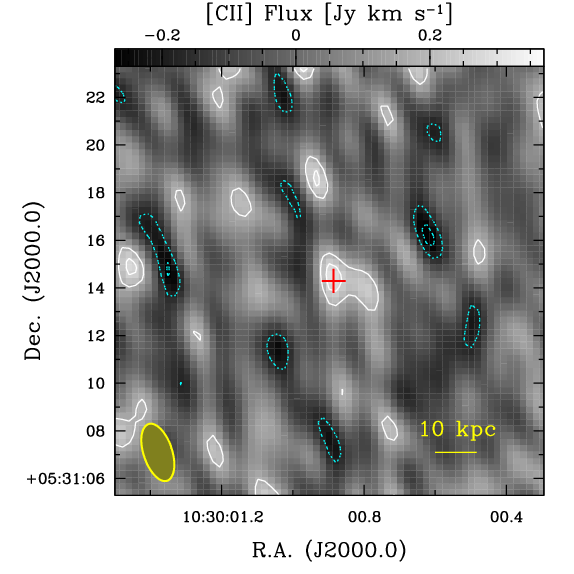}
 \includegraphics[width=0.33\textwidth]{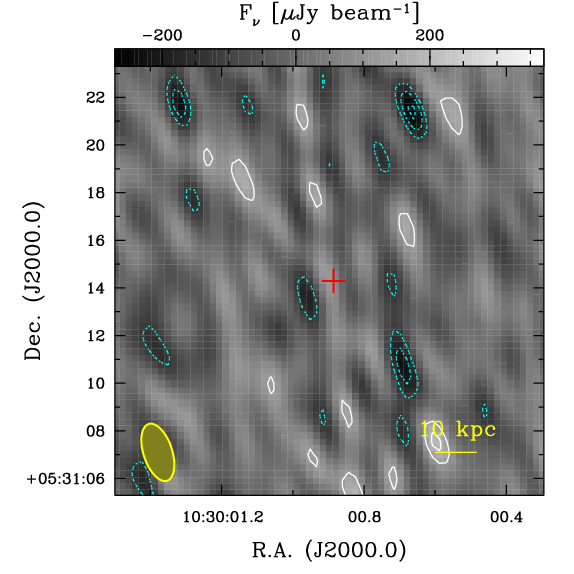}
\end{center}
 \caption{NOEMA observations of ID22914. {\em Left:} Single-pixel extraction of the 1.2\,mm spectrum of ID22914. The green dotted band gives the 1$\sigma$ range of each channel. The vertical bars show the inferred \Cii{} redshift of ID22914 (blue long-dashed line) and of J1030+0524 (red short-dashed line). {\em Middle:} Moment zero map of the \Cii{} line in ID22914. The field is $19''\times19''$, corresponding to the size of the primary beam. The position of the optical counterpart, located at the pointing center, is indicated by a red cross. The beam and the scale size in physical units are also reported. The contours give the $\pm$2, 3, 4, 5$\sigma$ levels, with $\sigma$=87\,mJy\,\kms{}\,beam$^{-1}$. Positive and  negative contours are shown in solid white and dotted cyan lines, respectively. ID22914 is detected at $\sim3.5$$\sigma$ level. {\em Right:} Continuum map of ID22914. The symbols are the same as in the central map, with positive and negative contours indicating  the $\pm$2, 3, 4, 5$\sigma$ levels and $\sigma$=62.3\,$\mu$Jy\,beam$^{-1}$.  No source is detected. }
 \label{fig_cii}
\end{figure*}

\subsection{NOEMA 1.2mm observations}

We observed ID22914 with NOEMA in two executions on 2018, January 13 and 28, with the array in compact (9D) configuration, as part of the program W17EW (PI: Decarli). The baseline range was 15--180\,m. The program was executed in typical winter weather conditions, with system temperature $T_{\rm sys}=110-145$\,K ($T_{\rm sys}=200-250$\,K) and a precipitable water vapor column of $\sim1$\,mm ($\sim2$\,mm) on January 28 (13). We took advantage of the new PolyFix correlator to simultaneously collect a total bandwidth of 15.6 GHz split in an upper and lower side band, encompassing the 270.4--278.2 GHz and 254.8--262.6 GHz bands, respectively. Quasars 1055+018 and J1018+055 served as phase and amplitude calibrators, while a set of millimeter-bright sources (LKHA101, 0851+202, 3C273, Vesta, MWC349) served as bandwidth and flux calibrators. We reduced the data using the January 2019 version of \textsf{clic}, in the \textsf{GILDAS} suite. The residual rms of the phase calibration was $\lsim20$ deg for most of baselines. Amplitude residual rms was $<10$\% in all cases. The final ($u$,$v$) table comprises 9330 visibilities, corresponding to 3.24\,hr of integration on source (nine-antenna equivalent). 

We imaged the visibilities using the software \textsf{mapping} in the \textsf{GILDAS} suite. At the tuning frequency of 260.3 GHz, the half power primary beam width is $19.4''$. We adopted natural weighting, yielding a synthesized beam of $2.5''\times1.2''$ with PA=18$^\circ$. The elongated beam is a direct consequence of the relatively short track and the equatorial position on sky of the targeted field. In the creation of the imaged cube, we re-binned the spectral axis into 50\,\kms{} wide channels ($\sim43$\,MHz at 260 GHz). The typical rms per channel is 0.91\,mJy\,beam$^{-1}$ in the lower side band, and 1.12\,mJy\,beam$^{-1}$ in the upper side band. We also created a collapsed continuum image, capitalizing on the whole available bandwidth. This 1.2\,mm continuum image reaches an rms of 62.3\,$\mu$Jy\,beam$^{-1}$.

\section{Results}\label{sec_results}

\subsection{Optical data}

The Keck spectrum of ID22914 shows a blue slope of the continuum with a sharp break associated with the Gunn--Peterson absorption (see Fig.~\ref{fig_lya}). The rest-frame UV color is consistent with the lack of a detection in the Y ($>$24.79) and J ($>$25.00) bands (limits reported at   3$\sigma$ significance). We measured the redshift by cross-correlating the DEIMOS spectrum with a Lyman-break galaxy template which incorporates the expected Ly$\alpha$ decrement at these redshifts. The template was taken from \citet{talia12}, and corrected for the Gunn--Peterson absorption expected at $z\approx 6.3$. The  best solution is {\it z}$\,\approx\,$6.3, thus matching the redshift of the quasar J1030+0524. However, the modest signal-to-noise ratio of the spectrum, 
the lack of a clear Ly$\alpha$ emission line, 
and the well-known fact that the Gunn--Peterson dropout associated with the hydrogen Ly$\alpha$ line can be off by hundreds of \kms{} (and up to a thousand) with respect to the systemic redshift \citep[see, e.g.,][]{venemans16}, 
all make it hard to accurately measure the redshift of this source via optical spectroscopy.
The resulting redshift uncertainty ($\Delta${\it z}$\,\approx\,$0.1) is insufficient in order to estimate the quasar--galaxy separation along the line of sight.

\subsection{Millimeter data}

Our NOEMA observations are sensitive to \Cii{} emission in the ranges $5.832<z<6.029$ and $6.237<z<6.459$. At the position of ID22914, the most significant feature in the whole range is a tentative emission line at 259.7 GHz (see Fig.~\ref{fig_cii}). We fit the line with a Gaussian profile, using our custom  Markov chain Monte Carlo code \textsf{smc}. The line has an integrated flux of  $0.37_{-0.06}^{+0.11}$ Jy\,\kms{}, and a width of $220_{-75}^{+90}$\,\kms. We created a moment-zero map of \Cii{} by integrating over the line width. This is shown in Fig.~\ref{fig_cii}, along with the map of the dust continuum. From the line moment-zero map, we infer a detection significance of 3.5$\sigma$. Assuming that the line is identified as \Cii{} (the brightest emission line that we expect at these frequencies, for a \zgt6 source), the observed line frequency implies a \Cii{} redshift of $z_{\rm [CII]}=6.3186_{-0.0006}^{+0.0009}$, in good agreement with the Ly$\alpha$-based estimate from our Keck spectrum. At this redshift, the associated line luminosity is $L_{\rm [CII]}=3.0\times10^8$\,\Lsun{}. The continuum emission is  undetected in ID22914. At 3$\sigma$ this implies a 1.2\,mm flux density $<189$\,$\mu$Jy\,beam$^{-1}$. In order to infer integrated infrared luminosities in the absence of a good sampling of the dust spectral energy distribution, it is common practice to assume that the dust emission is described by a modified blackbody  with a given emissivity index $\beta$ and dust temperature $T_{\rm dust}$. High-$z$ quasars typically have $T_{\rm dust}\approx 50$\,K \citep[e.g.,][]{beelen06,leipski14}. The cosmic infrared background sets a lower limit to $T_{\rm dust}>18.3$\,K at $z$=6.3. As ID22914 is not as extreme a system as quasar host galaxies at similar redshifts, it is reasonable to assume an intermediate dust temperature $T_{\rm dust}$=35\,K, comparable with typical values observed in intermediate-redshift main sequence galaxies \citep[e.g.,][]{magnelli14}. For the emissivity index, we adopt $\beta$=1.6 from \citet{beelen06}, although we note that the accurate choice has minimal impact on the integrated infrared luminosity (e.g., the estimated infrared luminosity only changes by a factor of $\approx 2\times$ for $\beta$ ranging between 1 and 2). By integrating the template in the rest-frame 8--1000\,$\mu$m range, we infer an infrared luminosity $L_{\rm IR}<2.2\times10^{11}$\,\Lsun{} (at 3$\sigma$), after correcting for the effects of the cosmic microwave background \citep{dacunha13}.

This value corresponds to a SFR$_{\rm IR}<33$\,\Msun{}\,yr$^{-1}$, following \citet{kennicutt12}, which  suggests that ID22914 is not a very dusty galaxy, in agreement with the blue rest-frame UV continuum. The implied obscured-to-unobscured SFR ratio is $\lsim$2. We note that all of these limits would be three times higher if we assumed $T_{\rm dust}$=50\,K, thus leaving room for a larger contribution of the obscured component to the star formation budget. In this case, the blue UV spectrum might be explained by a patchy geometry of the dust reddening.

We ran a blind search for emission lines at any position and frequency in the cubes. We used the \textsf{findclumps} software \citep{walter16,decarli19}. In brief, the code runs a floating average of channels over a wide range of line kernel widths, and searches for high-significance peaks in each collapsed channel. By comparing the statistics of positive and negative line candidates, we assessed the probability that a line is real (under the assumption that the noise is symmetric around zero, and that positive peaks are a mixed bag of real lines and noise peaks, while negative peaks are purely due to noise). Only two line candidates were found at S/N$>$5  (suggesting high reliability) in the cube within the primary beam, both at $\sim 259.2$\,GHz. A visual inspection of the maps suggests however that these lines might be associated with an imaging artifact, possibly related to the limited ($u$,$v$) sampling. 

\begin{figure}
\begin{center}
 \includegraphics[width=0.49\textwidth]{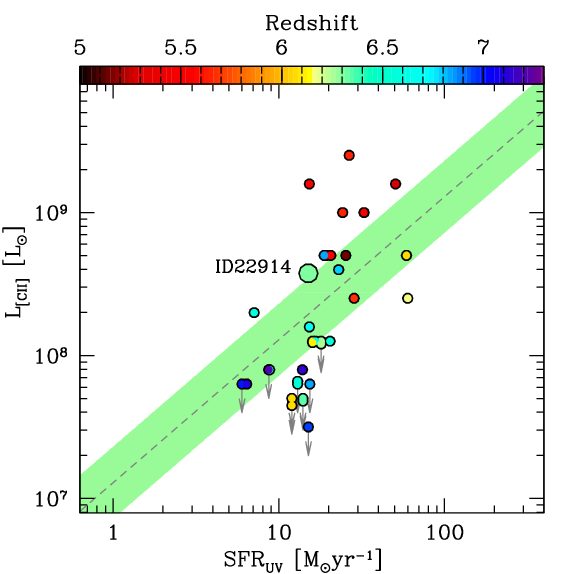}
\end{center}
 \caption{Luminosity of \Cii{}  as a function of UV-based SFR in ID22914 (large symbol) and in a compilation of UV-selected star-forming  galaxies at $z>5$ \citep[see][and references therein]{carniani18}. The relation for intermediate- to low-redshift main sequence star-forming galaxies from \citet{herreracamus18} is also shown for  comparison as a dashed line, with its 1$\sigma$ scatter  in green shading. ID22914 falls close to the expected relation, and within the scatter of values observed in similar galaxies at high redshift.}
 \label{fig_sfr_cii}
\end{figure}


\section{Discussion and conclusions}

The Keck spectrum of ID22914 provides a strong prior on the redshift of the source, which strengthens the significance of the \Cii{} line detection. The line luminosity matches our expectations for  a galaxy of this type: In Fig.~\ref{fig_sfr_cii}, we compare the observed \Cii{} luminosity and the UV-based SFR  in ID22914 with what is observed in the compilation of UV-selected $z$=5--7 galaxies in \citet{carniani18}. We find that ID22914 has a \Cii{} luminosity comparable with other galaxies at a similar SFR when we consider global source estimates (i.e., when we do not split galaxies into individual components). We also do not include high-$z$ sub-millimeter galaxies and quasars observed in \Cii{} in our comparison as for such sources it is impossible to estimate the UV-based SFR. The \Cii{} luminosity in ID22914 is lower than the typical values observed in the $z\sim5.5$ sample from \citet{capak15}, which comprises the most massive and dustiest sources in the \citet{carniani18} sample, but it is in excellent agreement with the expectations for low- to intermediate-redshift main sequence galaxies in the SHINING survey \citep{herreracamus18}. 

The redshift difference between ID22914 and J1030+0524 is $\Delta z=0.010$, corresponding to 3.9 comoving Mpc along the line of sight (assuming that the difference is solely attributed to the Hubble flow). The relative line-of-sight velocity difference, $\Delta v$=$c\,\Delta z/(1+z)$, is only 410\,\kms{}. Such a small redshift and velocity difference strongly points to the two galaxies belonging to a common large-scale structure that  extends over several physical megaparsecs.

The success of this pilot experiment on a single galaxy corroborates the case for (sub-)millimeter spectroscopy in order to accurately pin down the redshift of faint high-redshift galaxies, along the lines of successful searches for \Cii{} and \Oiii{} emission in Lyman-break galaxies at the highest redshifts \citep[see, e.g.,][]{inoue16,bradac17,laporte17,hashimoto18,hashimoto19,tamura19}. This holds valid also beyond the ``tip of the iceberg'' of dusty, highly star-forming galaxies. The upgraded capabilities offered by IRAM / NOEMA now enable sensitive \Cii{} investigations of typical galaxies at cosmic dawn within a few hours of integration, thus making the study of samples of these galaxies accessible to accuracy levels that are intrinsically not achievable with tens of hours of integration with state-of-the-art optical or near-infrared spectrographs. This approach can also be used in combination with ALMA in order to expand similar studies to the southern sky and to extend the study to higher frequency lines. For example, at $z\approx6.3$, the \Oiii{} 88\,$\mu$m is shifted to $\nu_{\rm obs}=465$\,GHz, where it can be secured with ALMA band\,8 observations (atmospheric transparency  up to $\sim45$\% for a precipitable water vapor of 1\,mm). A quantitative assessment of the galactic overdensities around \zgt6 quasars is thus now in reach.

\begin{acknowledgements}
We thank the anonymous referee for a prompt and helpful feedback that strengthened our manuscript. RD thanks Nichol Cunningham, Charl\`{e}ne Lefevre, Cinthya Herrera for support with the data handling, and acknowledges the hospitality of IRAM during his data reduction visit in January 2019. 
Based on observations carried out with the IRAM Interferometer NOEMA. IRAM is supported by INSU/CNRS (France), MPG (Germany) and IGN (Spain). This publication has received funding from the European Union’s Horizon 2020 research and innovation programme under grant agreement No 730562 [RadioNet].
\end{acknowledgements}

\label{lastpage}

\end{document}